\documentclass[a4paper,fleqn,usenatbib]{mnras}

\usepackage[T1]{fontenc}
\usepackage{ae,aecompl}

\usepackage{todonotes}
\newcommand{\pn}{{\text{\sc pn}}}
\newcommand{\mos}[1]{{\text{\sc mos#1}}}
\newcommand{\draco}{{\text{Dra}}}

\usepackage{geometry,fullpage,bm}
\usepackage[font=footnotesize]{caption}
\usepackage{lineno}

\newcommand{\flux}{\,\unit{cts/sec/cm^2}}

\usepackage{amssymb,amsfonts,amsmath,paralist,xspace}

\usepackage{etoolbox} %
\preto\tabular{\setcounter{magicrownumbers}{0}}%
\newcounter{magicrownumbers}%
\newcommand\rownumber{\stepcounter{magicrownumbers}\arabic{magicrownumbers}}%

\usepackage{longtable}
\usepackage[normalem]{ulem} 
\usepackage{relsize,color,xcolor} 
\usepackage{booktabs} 
\newcommand {\otoprule }{\midrule [\heavyrulewidth]} 

\usepackage{units}

\usepackage{grffile} 
\usepackage{graphicx} 
\graphicspath{{./figures/}}

\IfFileExists{srcltx.sty}{\usepackage[active]{srcltx}}

\newcommand{\kev}{\:\mathrm{keV}}

 \newcommand{\be}
{\begin{equation}} \newcommand{\ee} {\end{equation}}   

\def\be{\begin{eqnarray}} \def\ee{\end{eqnarray}} 
 \renewcommand{\S}{\mathcal{S}} 
\newcommand{\dm}{{\textsc{dm}}} 
 
\newcommand{\xmm}{\textit{XMM-Newton}\xspace}
 
\newcommand{\suza}{\textsl{Suzaku}\xspace} 
\newcommand{\fov}{\mathrm{fov}}

\renewcommand{\S}{\mathcal{S}}

\title[Searching for decaying dark matter in Draco]
{Searching for decaying dark matter in deep \xmm\ observation of the Draco dwarf spheroidal}

\author[O.~Ruchayskiy et al.]
{Oleg Ruchayskiy,$^{1,2}$\thanks{E-mail: oleg.ruchayskiy@epfl.ch} Alexey Boyarsky,$^3$ Dmytro Iakubovskyi,$^{2,4}$
\newauthor Esra Bulbul,$^5$ Dominique Eckert,$^6$ Jeroen Franse,$^{3,7}$ Denys Malyshev,$^6$
\newauthor Maxim Markevitch,$^8$ Andrii Neronov$^6$ \\
\\
$^1$Ecole Polytechnique F\'ed\'erale de Lausanne, FSB/ITP/LPPC, BSP 720, CH-1015, Lausanne, Switzerland\\
$^2$Niels Bohr Institute \& Discovery Center, Blegdamsvej 17, DK-2100 Copenhagen, Denmark\\
$^3$Instituut-Lorentz for Theoretical Physics, Universiteit Leiden, Niels Bohrweg 2, Leiden, The Netherlands\\
$^4$Bogolyubov Institute of Theoretical Physics, Metrologichna Str. 14-b, 03680, Kyiv, Ukraine\\
$^5$Kavli Institute for Astrophysics and Space Research, Massachusetts Institute of Technology, 77 Massachusetts Avenue, \\
Cambridge, MA 02139\\
$^6$Department of Astronomy, University of Geneva, ch.\ d'Ecogia 16, CH-1290 Versoix, Switzerland\\
$^7$Leiden Observatory, Leiden University, Niels Bohrweg 2, Leiden, The Netherlands\\
$^8$NASA Goddard Space Flight Center, Code 662, 8800 Greenbelt Rd., Greenbelt, MD 20771, USA 
}
\date{Accepted XXX. Received YYY; in original form ZZZ}

\pubyear{2016}

\begin{document}
\label{firstpage}
\pagerange{\pageref{firstpage}--\pageref{lastpage}}
\maketitle

\begin{abstract}
We present results of a search for the 3.5 keV emission line in our recent
very long ($\sim$ 1.4 Ms) \xmm\ observation of the Draco dwarf
spheroidal galaxy. The astrophysical X-ray emission from such dark
matter-dominated galaxies is faint, thus they provide a test for the dark
matter origin of the 3.5 keV line previously detected in other massive, but
X-ray bright objects, such as galaxies and galaxy clusters. We do not detect
a statistically significant emission line from Draco; this constrains the
lifetime of a decaying dark matter particle to $\tau>(7-9)\times 10^{27}$ s
at 95\% CL (combining all three \xmm\ cameras; the interval corresponds to the
uncertainty of the dark matter column density in the direction of Draco). The PN camera,
which has the highest sensitivity of the three, does show a positive
spectral residual (above the carefully modeled continuum) at $E=3.54\pm0.06$
keV with a $2.3\sigma$ significance. The two MOS cameras show
less-significant or no positive deviations, consistently within $1\sigma$
with PN. Our Draco limit on $\tau$ is consistent with previous detections in
the stacked galaxy clusters, M31 and the Galactic Center within their
$1-2\sigma$ uncertainties, but is inconsistent with the high signal
from the core
of the Perseus cluster (which has itself been inconsistent with the rest of the
detections). We conclude that this Draco observation does not exclude the
dark matter interpretation of the 3.5 keV line in those objects.
\end{abstract}


\section{Introduction}
\label{sec:introduction}

An emission line-like spectral feature at energy $E\sim 3.5$~keV has recently been observed in the
long-exposure X-ray observations of a number of dark matter-dominated objects:
in a stack of 73 galaxy clusters \citep{Bulbul:14a} and in the Andromeda galaxy
and the Perseus galaxy cluster \citep{Boyarsky:14a}. 
The possibility that this spectral feature may be the signal from decaying
dark matter has sparked a lot of interest in the community, and many dark
matter models explaining this signal have been proposed (see e.g.\
\cite{Iakubovskyi:14} and refs.\ therein). The signal was  subsequently
detected 
in the Galactic Center
\citep{Riemer-Sorensen:14,Boyarsky:14b,Jeltema:14a,Carlson:14},
in the center of the Perseus galaxy cluster with \suza~\citep{Urban:14}, in a stacked spectrum of
a new set of galaxy clusters~\citep{Iakubovskyi:15b}, but not found in stacked
spectra of dwarf spheroidal galaxies~\citep{Malyshev:14}, in outskirts of
galaxies~\citep{Anderson:14}, in the diffuse X-ray background~\citep{Figueroa-Feliciano:15, Sekiya:15}.

There are three classes of non-dark matter explanations: a statistical
fluctuation, an unknown systematic effect or an atomic line. Instrumental
origins of this signal have been shown to be unlikely for a variety of
reasons: the signal is present in the spectra of galaxy clusters in all of \xmm\ detectors and also in
\textit{Chandra}, yet it is absent in a very-long exposure \xmm~\citep{Boyarsky:14a} or
\suza~\citep{Sekiya:15} blank sky backgrounds. 
The
position of the line in galaxy clusters scales correctly with 
redshift~\citep{Bulbul:14a,Boyarsky:14a,Iakubovskyi:15b}, 
and the line has radial surface brightness profiles in the
Perseus cluster (except its core~\citep{Bulbul:14a}) and Andromeda galaxy~\citep{Boyarsky:14a} consistent
with our expectations for decaying dark
matter and with the mass distribution in these objects.

The astrophysical explanation of this signal (e.g., an anomalously bright K {\sc xviii} 
line~\citep{Bulbul:14a,Riemer-Sorensen:14,Boyarsky:14b,Jeltema:14a,Carlson:14,Iakubovskyi:15b} 
or Ar {\sc xvii} satellite line~\citep{Bulbul:14a}, 
or a Sulfur charge exchange line~\citep{Gu:15}) require a significant stretch of the astrophysical 
emission models, though they can be unambiguously tested only with the high spectral resolution of the
forthcoming {\em Astro-H} and 
{\em Micro-X}~\citep{Mitsuda:14,Koyama:14,Kitayama:14,Figueroa-Feliciano:15,Iakubovskyi:15a}
microcalorimeters.

The dark matter interpretation of the origin of the line  allows for 
non-trivial consistency check by comparing observations of different objects:
the intensity of the line should correlate with the \emph{dark matter column
  density} -- a quantity that is bracketed between roughly
$10^2\,M_\odot/\unit{pc}^2$ and $\text{few}\times 10^3\,M_\odot/\unit{pc}^2$ for all objects
from smallest galaxies for larger clusters~\citep{Boyarsky:09c,Boyarsky:09b}.
For example, the observation of the $3.5$~keV line in
M31 and the Perseus cluster puts a \emph{lower limit on the flux} expected from the
Galactic Center (GC). On the other hand, the non-detection of any signal in the off-center
observations of the Milky Way halo (the blank sky dataset of
\citep{Boyarsky:14a}) \emph{provides an upper limit} on the possible flux from
the GC, given the observational constraints on the DM distribution in the
Galaxy.  \cite{Boyarsky:14b} demonstrated the flux of the $3.5$~keV line
detected in the GC, falls into this
range.

To study this question further, we observed the central $r=14'$ of the Draco dwarf spheroidal galaxy with
\xmm\ with a very deep exposure of 1.4~Msec.
Dwarf spheroidal galaxies (dSphs) are the most extreme dark matter-dominated
objects known.  The observations of thousands of stars in the ``classical''
dwarf satellites of the Milky Way make possible the determination of their DM
content with very low uncertainties, with the Draco dSph being one of the best
studied (see~\cite{Geringer-Sameth:14} for the latest mass modeling).  
The relatively small uncertainty on the dark matter
column density in Draco and its ``faintness'' in X-rays due to lack of
gas or X-ray binaries allows us to devise a test of the decaying dark
matter interpretation of the $\sim$3.5~keV line, as we have a clear
prediction of the expected line flux for Draco based on the masses of
this galaxy and the other objects.

A clear detection of the $\sim 3.5$~keV line in Draco would
provide very convincing evidence for the decaying DM interpretation, as there
is no known physical process that would produce \emph{the same signal} over
such a broad range of objects and environments, with an intensity that scales
with the DM content, from galaxy clusters of huge masses and large abundances
of hot gas, through spiral galaxies, and then all the way down to dwarf
galaxies of very low magnitude and negligible gas content.

The estimated column density within the central $14'$ is a
factor of a few lower in dSphs (including Draco) than in the centers of nearby
spiral galaxies or clusters (albeit has a much lower uncertainty). Therefore,
to achieve the same signal-to-noise for a dSph as for a spiral galaxy for all
observationally possible ratios of the DM column density, one would need a
prohibitively long observation. The uncertainty in DM content of the galaxies
becomes crucial.  Therefore, the raw exposure time of the Draco observation (1.4~Msec)
has been chosen to match the shortest exposure expected to still allow for detection
of the weakest possible line (compatible with previous observations).
  
In this paper we describe the results of the analysis of these Draco
observations. 
  We do find weak positive residuals at the predicted energy above the
(carefully-modeled) continuum in the PN spectrum and in one of the two
MOS spectra, but at a low statistical significance that allows only an
upper limit on the line flux to be set. The upper limit is consistent with most of the
previous positive line detections, thus, we \emph{cannot exclude dark matter
decay} as origin of the $3.5$~keV line.

\begin{figure}
  \centering
  \includegraphics[width=0.99\linewidth]{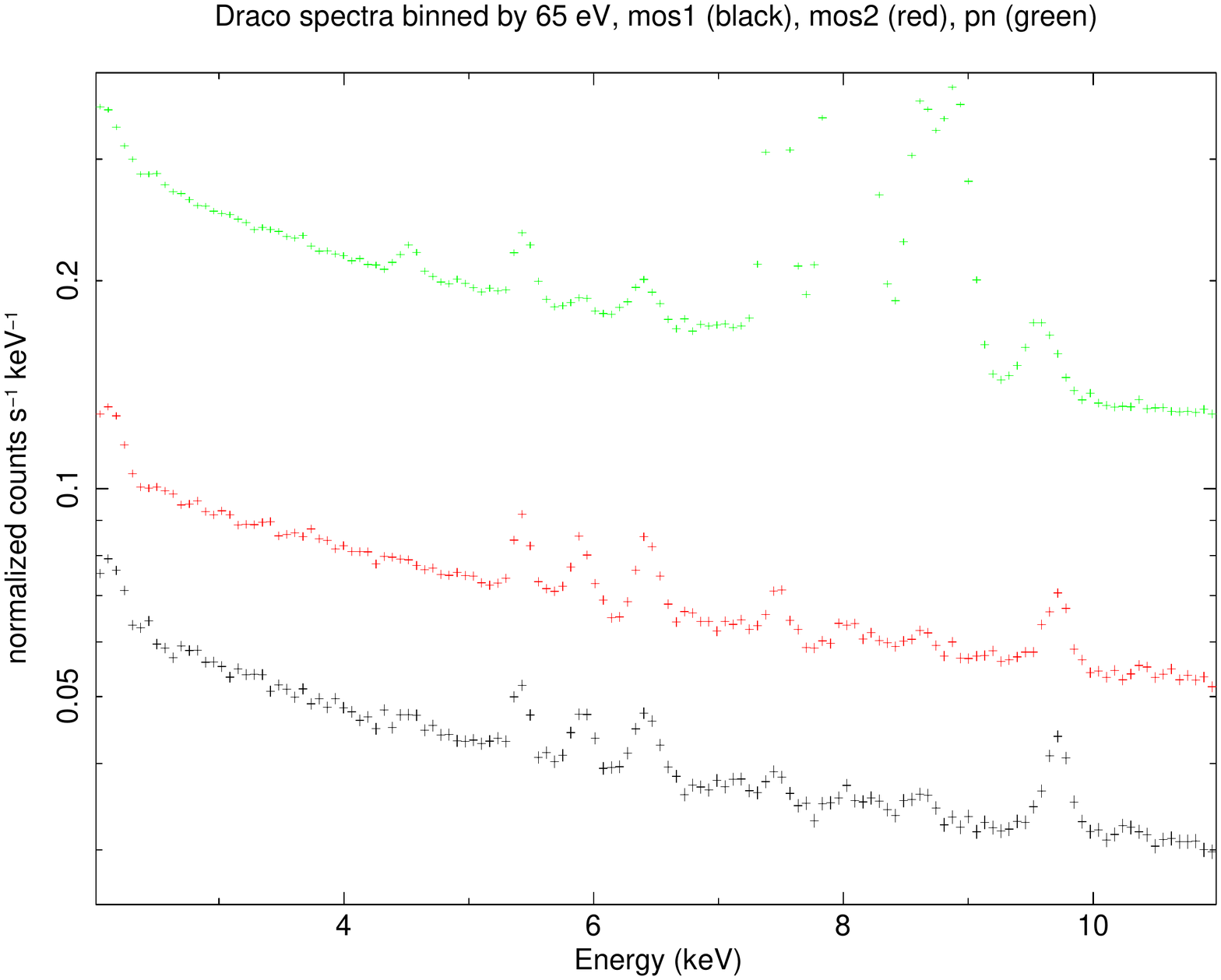}
  \caption{Spectra of Draco dwarf spheroidal seen by MOS1 (black), MOS2 (red) and PN (green) cameras.}
  \label{fig:spectra}
\end{figure} 

\begin{table*}
\centering
\begin{tabular}[c]{rlccc}
\hline
& ObsID~~~ & Observation date & Cleaned exposure [ksec] & Cleaned FoV [arcmin$^2$] \\
          &                  & MOS1/MOS2/PN          & MOS1/MOS2/PN \\
\hline
\rownumber & \texttt{0764800101} & 2015-03-18 & 31.6/36.0/12.4 & 320.7/573.7/553.1 \\
\rownumber & \texttt{0764800301} & 2015-03-26 & 23.9/28.2/13.5 & 319.4/575.2/549.8 \\
\rownumber & \texttt{0764800401} & 2015-03-28 & 41.3/42.0/30.5 & 316.8/571.6/545.3 \\
\rownumber & \texttt{0764800201} & 2015-04-05 & 26.1/27.6/17.2 & 315.4/571.5/549.0 \\
\rownumber & \texttt{0764800501} & 2015-04-07 & 47.5/49.5/24.7 & 314.6/567.1/543.4 \\
\rownumber & \texttt{0764800601} & 2015-04-09 & 52.5/52.0/38.9 & 314.2/568.5/542.9 \\
\rownumber & \texttt{0764800801} & 2015-04-19 & 25.4/29.3/12.8 & 320.3/573.7/554.8 \\
\rownumber & \texttt{0764800901} & 2015-04-25 & 36.0/44.2/16.5 & 318.3/574.3/550.7 \\
\rownumber & \texttt{0770180101} & 2015-04-27 & 34.1/35.9/20.9 & 323.8/579.5/548.2 \\
\rownumber & \texttt{0770180201} & 2015-05-25 & 51.9/53.4/32.4 & 314.2/569.4/541.2 \\
\rownumber & \texttt{0764800701} & 2015-06-15 & 54.2/54.5/47.7 & 312.9/566.3/530.0 \\
\rownumber & \texttt{0770180401} & 2015-06-18 & 50.5/50.1/40.3 & 315.4/564.1/536.2 \\
\rownumber & \texttt{0770180301} & 2015-07-01 & 52.5/54.8/47.4 & 311.4/565.9/535.2 \\
\rownumber & \texttt{0770180501} & 2015-07-31 & 49.1/50.2/41.0 & 315.6/571.0/538.9 \\
\rownumber & \texttt{0770180701} & 2015-08-22 & 38.0/38.5/25.2 & 320.1/572.8/546.8 \\
\rownumber & \texttt{0770180601} & 2015-09-01 & 46.6/49.0/26.2 & 318.5/573.9/548.7 \\
\rownumber & \texttt{0770180801} & 2015-09-03 & 64.2/65.8/48.0 & 322.1/577.9/537.7 \\
\rownumber & \texttt{0770190401} & 2015-09-11 & 50.1/50.0/40.2 & 328.2/583.9/565.8 \\
\rownumber & \texttt{0770190301} & 2015-09-21 & 22.5/24.8/11.6 & 324.1/578.6/554.0 \\
\rownumber & \texttt{0770190101} & 2015-09-23 & 18.1/19.5/0.9 & 334.1/590.5/566.5 \\
\rownumber & \texttt{0770190201} & 2015-09-25 & 18.8/20.1/9.3 & 322.3/579.5/553.9 \\
\rownumber & \texttt{0770190501} & 2015-10-11 & 30.8/31.5/20.3 & 320.2/576.8/542.9 \\
\rownumber & \texttt{0770180901} & 2015-10-13 & 19.2/20.8/11.4 & 324.6/579.1/560.9 \\
\rownumber & \texttt{0770190601} & 2015-10-15 & 7.6/11.2/4.6 & 329.1/584.0/553.8 \\
\rownumber & \texttt{0770190701} & 2015-10-17 & 43.7/45.0/35.5 & 320.5/572.8/537.4 \\
\rownumber & \texttt{0770190801} & 2015-10-19 & 31.7/32.5/22.4 & 326.2/583.2/547.2 \\
\hline
& Total               &         & 967.8/1016.1/651.8 & 318.9/573.5/543.9
\end{tabular}
\caption{Cleaned exposures and fields-of-view for 
 26 AO14 observations of Draco dSph used in our analysis,
 see Appendix~\protect\ref{sec:data-analysis} for details. 
 Notable difference between MOS1 and MOS2 fields of view is due to micrometeoroid damages in MOS1, 
see e.g.~\protect\cite{Abbey:06}.}
\label{tab:observations}
\end{table*}

\section{Results}\label{sec:results}

The data preparation and analysis is fairly standard and is described in Appendix~\ref{sec:data-analysis}. 
The extracted spectra are shown in
Fig.~\ref{fig:spectra}. 
They are dominated by the instrumental and Cosmic (CXB) X-ray backgrounds, which we will
carefully model below to see if there is any residual flux at 3.5~keV.
As has been stressed in our previous works, the 3.5~keV line feature
is so weak that the continuum in its spectral vicinity has to be
modeled to a very  high precision to be able to detect the line --
more precisely than what's acceptable in the usual X-ray observation.

The basic information about the spectra is listed in Table~\ref{tab:observations}.
As a baseline model, we used the combination of Cosmic X-ray Background 
(extragalatic \texttt{powerlaw} folded with the effective area of the instrument) 
and instrumental background (instrumental \texttt{powerlaw} not folded with instrument response, 
plus several narrow 
\texttt{gaussian}s describing fluorescence lines) components. The model parameters are summarized
in Table~\ref{tab:model-parameters}; they are consistent with previous measurements and are consistent among
MOS1, MOS2 and PN cameras. Particular differences of the models among individual cameras are described below.

\subsection{Line detection in PN camera}
\label{sec:pn-flux}

We obtained an excellent fit to the EPIC PN spectrum, with $\chi^2 = 47.0$ for 58 d.o.f. 
(Table~\ref{tab:model-parameters}). 
The spectrum shows a faint line-like residual
at the right energy, $E=3.54^{+0.06}_{-0.05}$~keV. Its flux is
\begin{equation}
  \label{eq:2}
  F_\pn = \left\{
    \begin{aligned}
   1.65^{+0.67}_{-0.70}\times 10^{-6} & \flux \\
  3.0^{+1.23}_{-1.29}\times 10^{-9}   & \flux/\unit{arcmin^2}    
    \end{aligned}\right.
\end{equation}

\begin{figure*}
  \centering
  \includegraphics[width=0.47\textwidth]{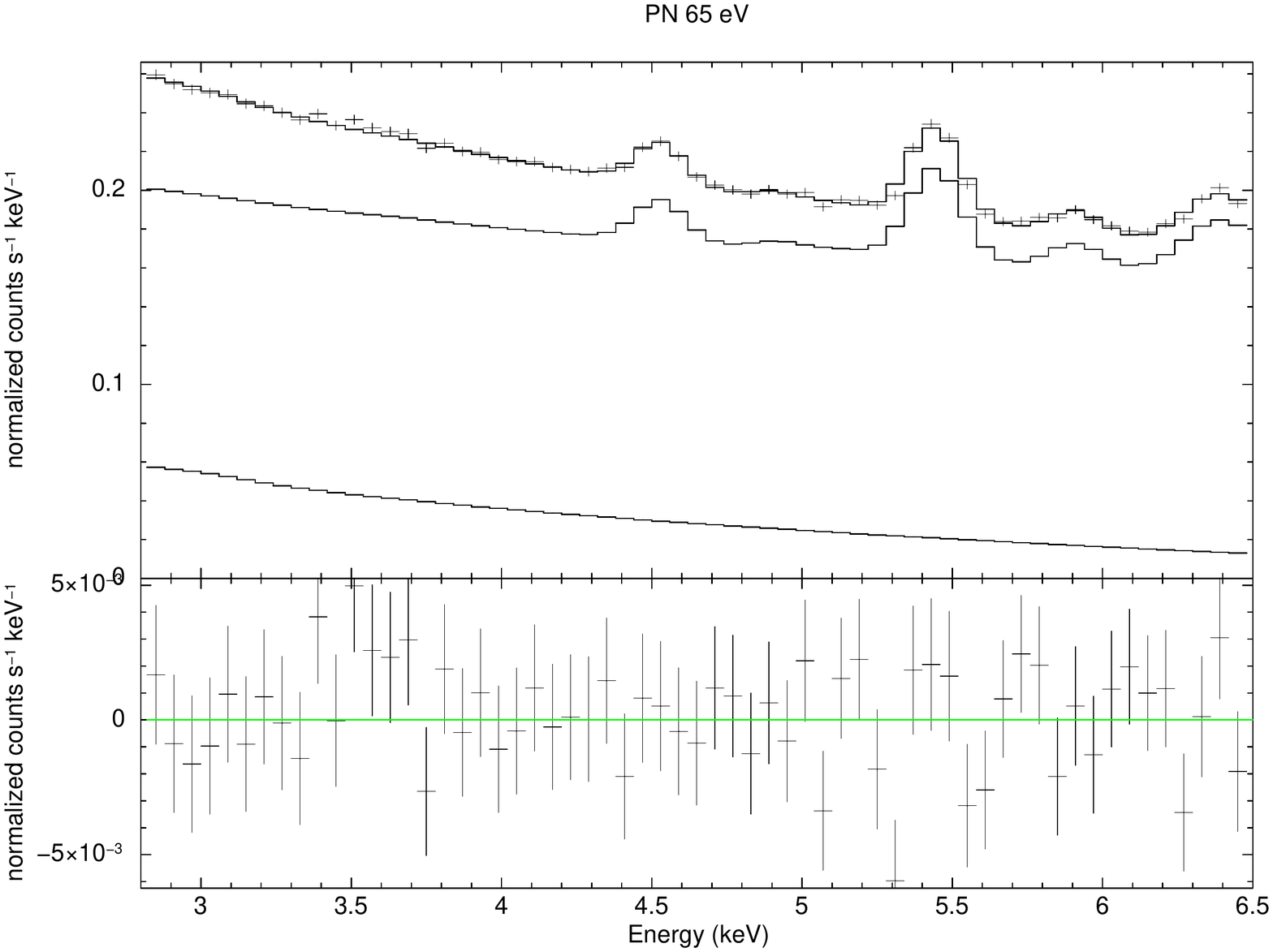}~
\includegraphics[width=0.53\textwidth]{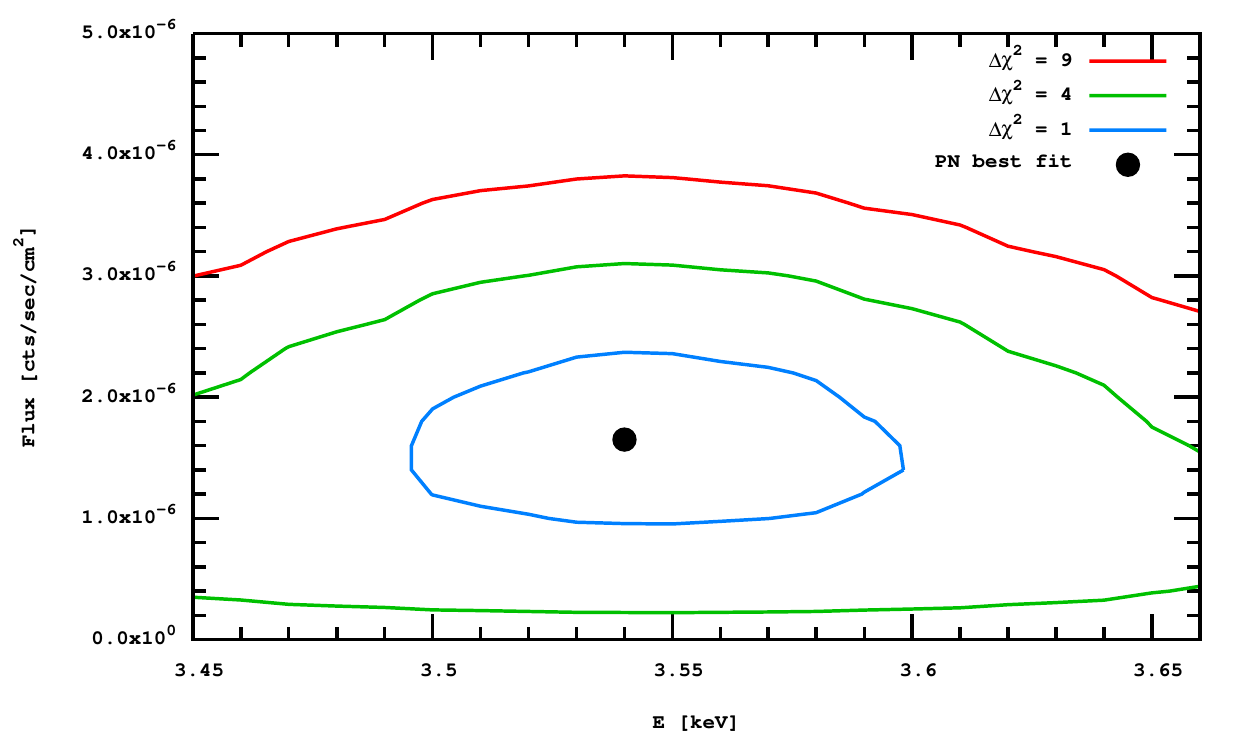}
\caption{Left panel: PN spectrum with \emph{unmodeled} feature at $3.54$~keV.
  Also shown instrumental (upper) and astrophysical (lower) components of the
  background model. Right panel: Best fit value (black square) and
  $\Delta \chi^2 = 1, 4, 9$ contours for the line in the PN camera.}
  \label{fig:pn_123sigma_predictions}
\end{figure*}

(where the bottom value corresponds to the surface brightness).
The improvement of fit when adding the line is $\Delta \chi^2 = 5.3$ for 2 additional d.o.f.. 
Thus, the detection has a relatively low significance of $2.3\sigma$.
The PN spectrum with the unmodeled line-like residual together with the $\Delta\chi^2 = 1,4,9$ contours is 
shown in Figure~\ref{fig:pn_123sigma_predictions}. In this Figure, we show only 2.8-6.5~keV range for clarity,
while the fit includes higher energies, as described in Appendix~\ref{sec:data-analysis}.

\subsection{MOS cameras}
\label{sec:rescaling-mos}

Next we turn to MOS1 and MOS2 cameras.  In MOS2 camera, we detected line-like
residuals at $\sim 3.38$~keV (significance $3.2\sigma$) and $\sim 3.77$~keV
(significance $3.5\sigma$), see
Table~\ref{tab:model-parameters} in Appendix~\ref{sec:data-analysis}.
The positions of these residuals are consistent
with K~K$\alpha$ and Ca~K$\alpha$ fluorescent lines, respectively. These
fluorescent lines have not been previously detected in the MOS cameras, but have
been detected in the PN camera with the enhanced calibration source (the
so-called \texttt{CalClosed} mode, see Fig.~6 of~\cite{Strueder:01} for
details). Another line-like residual at $\sim 3.05$~keV is visible (in the MOS2
spectrum only) at a 1.9$\sigma$ significance; we could not identify it,
though its energy is consistent with L and M lines of several heavy
metals.
In order to get as accurate a continuum model as
possible, we have included these weak detector lines as narrow
\texttt{gaussian} components in
our spectral model.

\begin{figure*}
  \centering
  \includegraphics[width=0.48\textwidth]{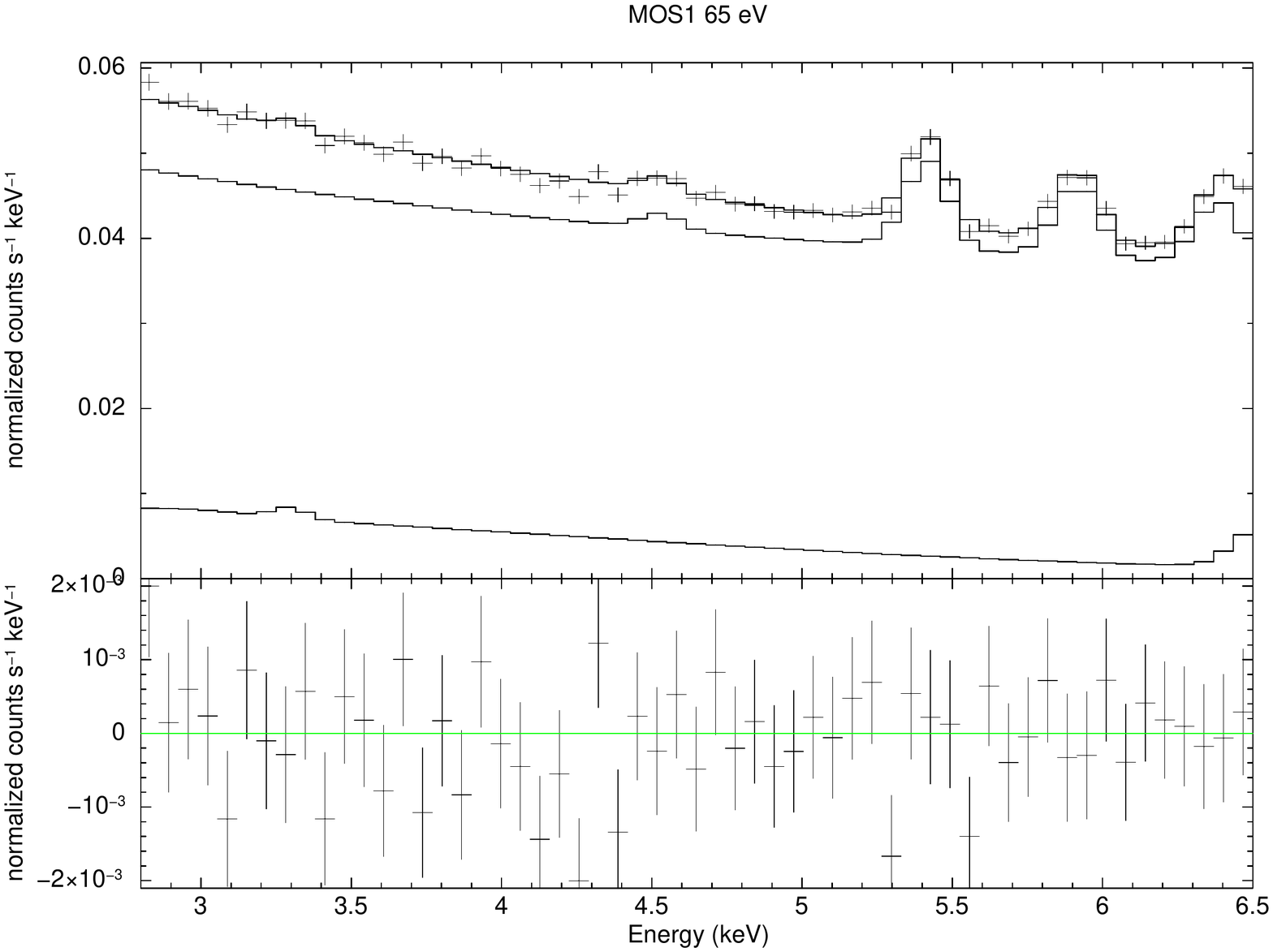}
  ~\includegraphics[width=0.48\textwidth]{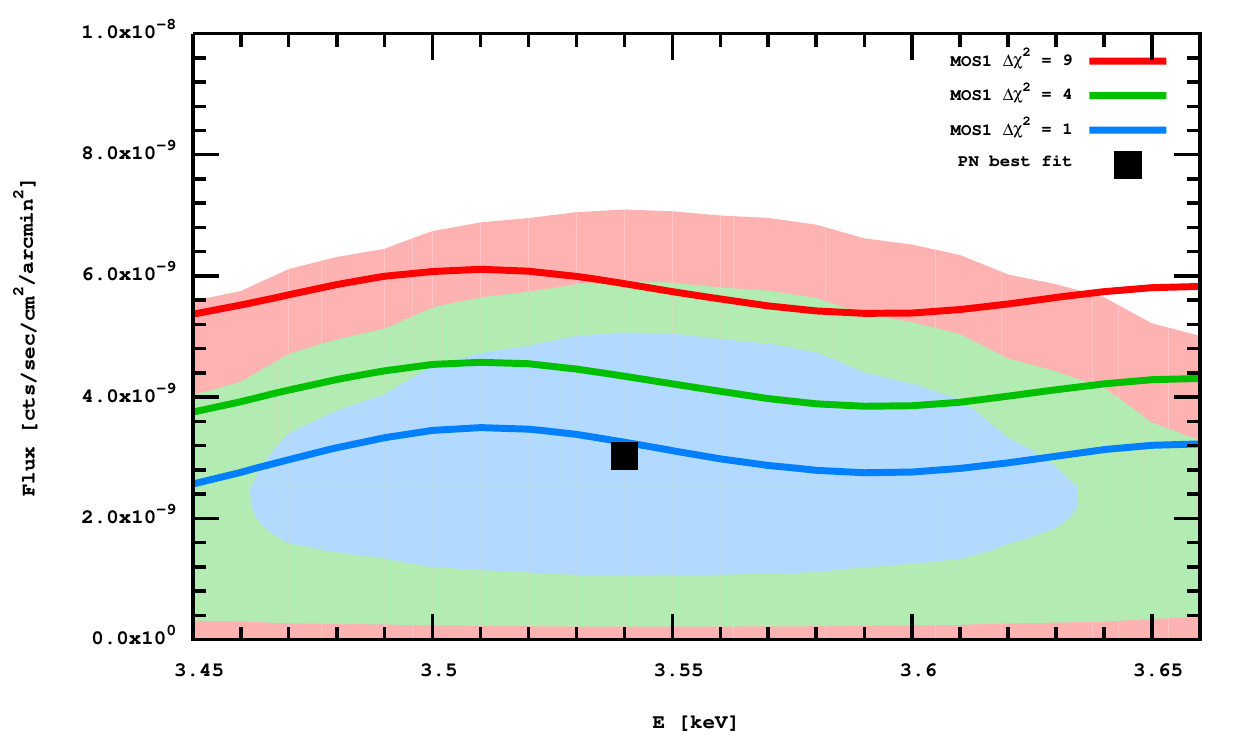}
  \caption{\textit{Left panel:} MOS1 spectrum and residuals. \textit{Right
    panel:} $\Delta \chi^2 = 1, 4, 9$ contours from the MOS1 camera (thick
    lines). The PN camera contours with $\Delta \chi^2 = 1, 4, 9$ are shown
    as shaded regions (identical to the contours in 
    Fig.~\protect\ref{fig:pn_123sigma_predictions}).}
  \label{fig:mos1-line-contours}
\end{figure*}

\begin{figure*}
  \centering
  \includegraphics[width=0.48\textwidth]{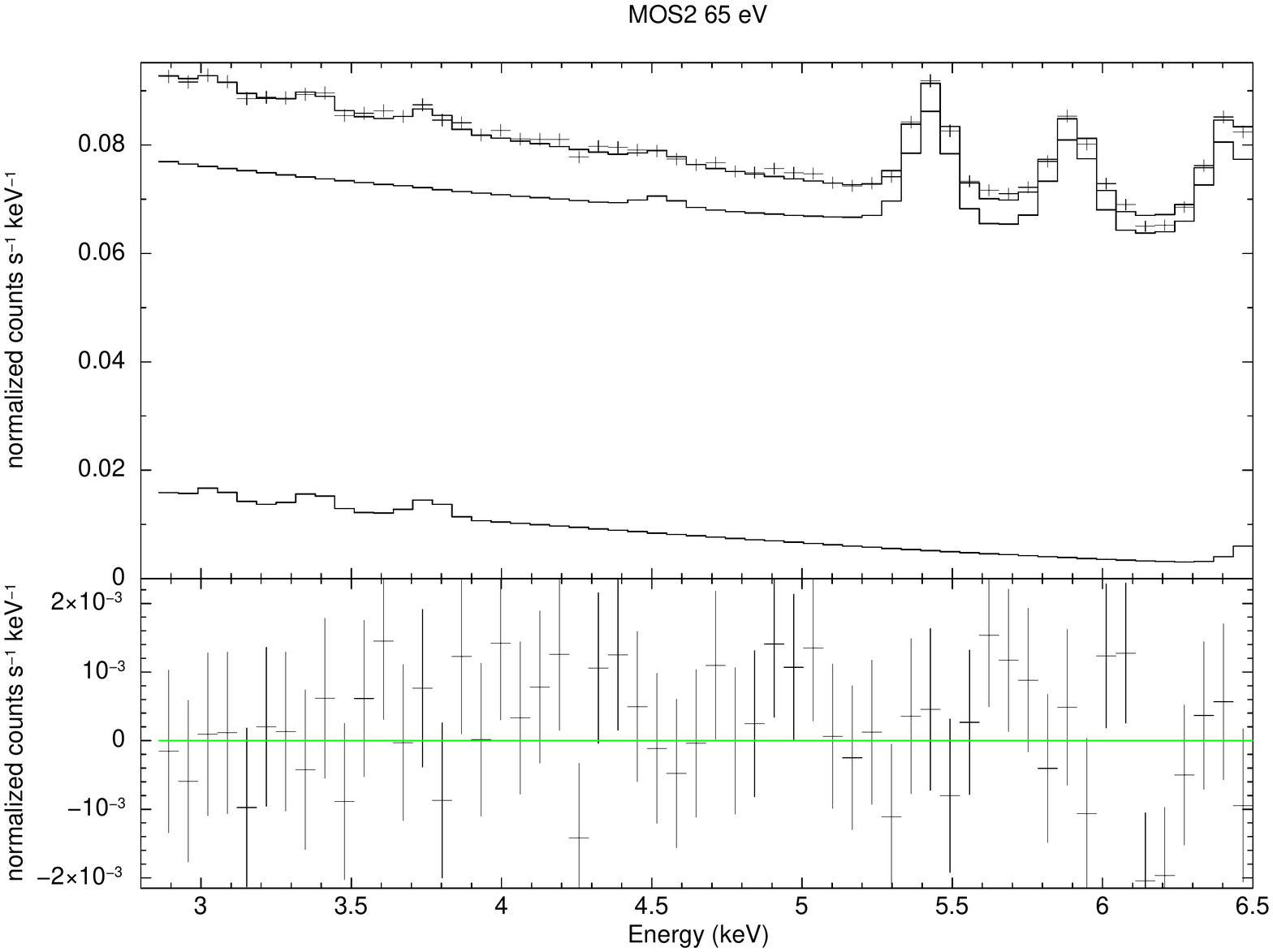}
  ~\includegraphics[width=0.48\textwidth]{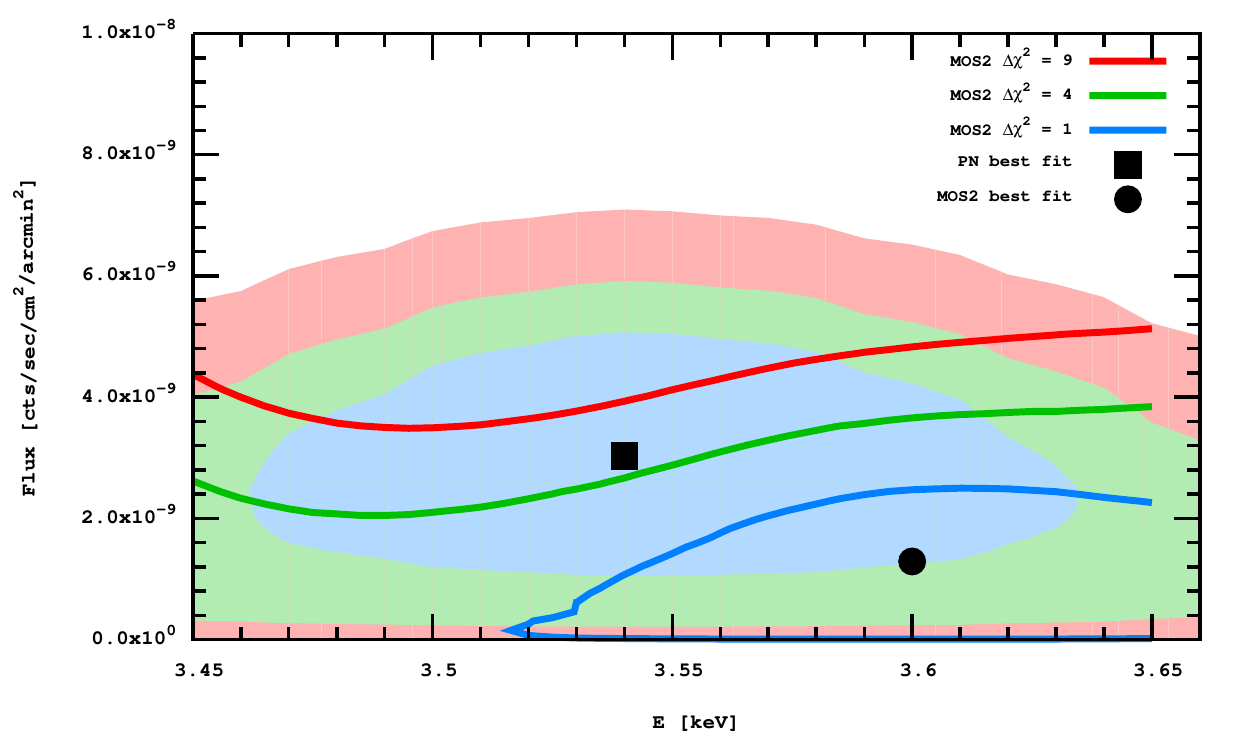}
  \caption{\textit{Left panel:} MOS2 spectrum with \emph{unmodeled} feature at
    $\sim$3.5~keV and residuals. \textit{Right panel:}
    $\Delta \chi^2 = 1, 4, 9$ contours from the MOS2 camera (thick lines) and
    the best fit value (black circle).  The PN camera contours with
    $\Delta \chi^2 = 1, 4, 9$ are shown as shaded regions
    (identical to the contours in Fig.~\protect\ref{fig:pn_123sigma_predictions}).}
  \label{fig:mos2-line-contours}
\end{figure*}

The MOS1 camera reveals no other lines in the region 3--4~keV, see left panel
in Figure~\ref{fig:mos1-line-contours}. 
The MOS2 camera has a hint of a residual ($\Delta \chi^2 = 1.2$) in the position $3.60\pm 0.07$~keV,
see left Figure~\ref{fig:mos2-line-contours}.
Assuming that the line detected in the PN camera is a physical line, we rescale the PN flux, given 
by Eq.~\eqref{eq:2}, according to the ratio of MOS1 and MOS2 FoV to that of the PN camera (see
Table~\ref{tab:observations}). 
Table~\ref{tab:cross-check-cameras} also shows a change in the $\chi^2$ (after running the new fit) 
when one adds a line with a fixed flux to the spectrum of MOS1 and MOS2 (allowing its
position to vary within $\pm 1 \sigma$ -- from $3.49$ to $3.60$~keV). 
Clearly the non-observation in MOS1 and MOS2
are consistent with PN observation at $\lesssim 1\sigma$ level.

\begin{table*}
  \centering
  \begin{small}
    \begin{tabular}{lccc}
      \toprule
      Camera                    & PN flux    & Predicted flux           & $\Delta\chi^2$ \\
                                & [$10^{-6} \flux$] & [$10^{-6} \flux$] &                \\
      \otoprule                                                         
      MOS1, PN best-fit         & 1.65              & 0.97              & 1.58           \\
      MOS2, PN best-fit         & 1.65              & 1.74              & 2.23           \\
      MOS1, PN $1\sigma$ lower  & 0.95              & 0.56              & 0.27           \\
      MOS2, PN $1\sigma$ lower  & 0.95              & 1.00              & 0.25           \\    
      \bottomrule
    \end{tabular}
  \end{small}
  \caption{Consistency check of MOS1 and MOS2 cameras with rescaled flux from PN camera. Line position is 
  allowed to vary within 1$\sigma$ bound for PN, i.e.\ 3.49--3.60~keV. See also Figures~\protect\ref{fig:mos1-line-contours}
  and~\protect\ref{fig:mos2-line-contours}.}
  \label{tab:cross-check-cameras}
\end{table*}

\subsection{Common fit of MOS1, MOS2 and PN cameras}\label{sec:common-fit}

Having shown that
the three XMM cameras are consistent with each other, we now perform a
common fit of all three. We kept the ratio of the \texttt{gaussian}
normalisations at $\sim 3.5$~keV fixed to
the ratios of the corresponding FoVs. 
This is justified if the surface brightness is uniform across 
the FoV of the XMM cameras. We estimated that assuming instead cuspy 
Navarro-Frenk-White dark matter profile, the expected surface brightness of 
3.5~keV features caused by decaying dark matter differs between
the cameras varies by no more than 15\%.
\footnote{While FoV of MOS1 is smaller than that of MOS2 by 44\%, 
this decrease is largely compensated by the fact that \emph{non-central} CCDs are shut down. 
So the observation of the central densest part of Draco is not affected.}
This different scaling of the signal between cameras would affect the results 
of our combined fit by less than 5\%.
The common fit finds a positive residual with
$\Delta\chi^2 = 2.9$ at $E =3.60\pm0.06$~keV
and the flux $F = 1.3^{+0.9}_{-0.7}\times 10^{-9}\flux/\unit{arcmin^2}$.  The
resulting flux is compatible with the best-fit PN flux
(Eq.~\eqref{eq:2}) at a $2\sigma$ level.

\begin{figure*}
  \centering
  \includegraphics[width=0.5\textwidth]{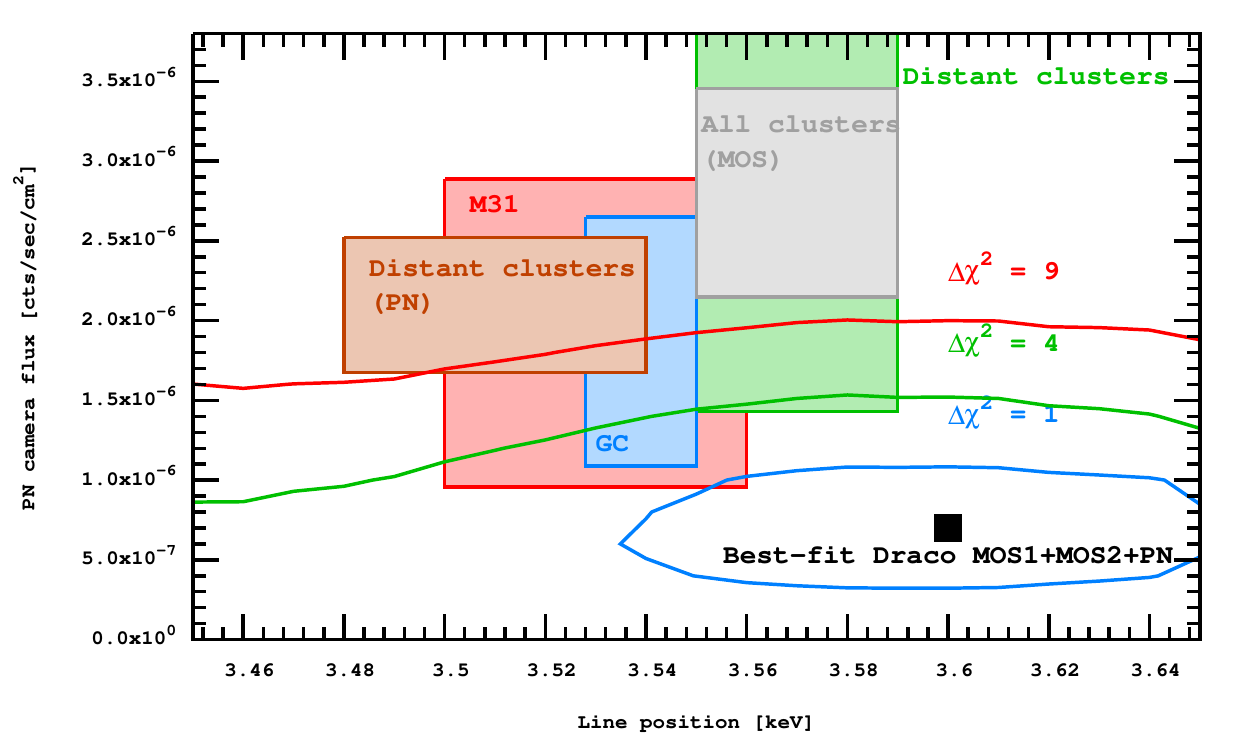}~%
  \includegraphics[width=0.5\textwidth]{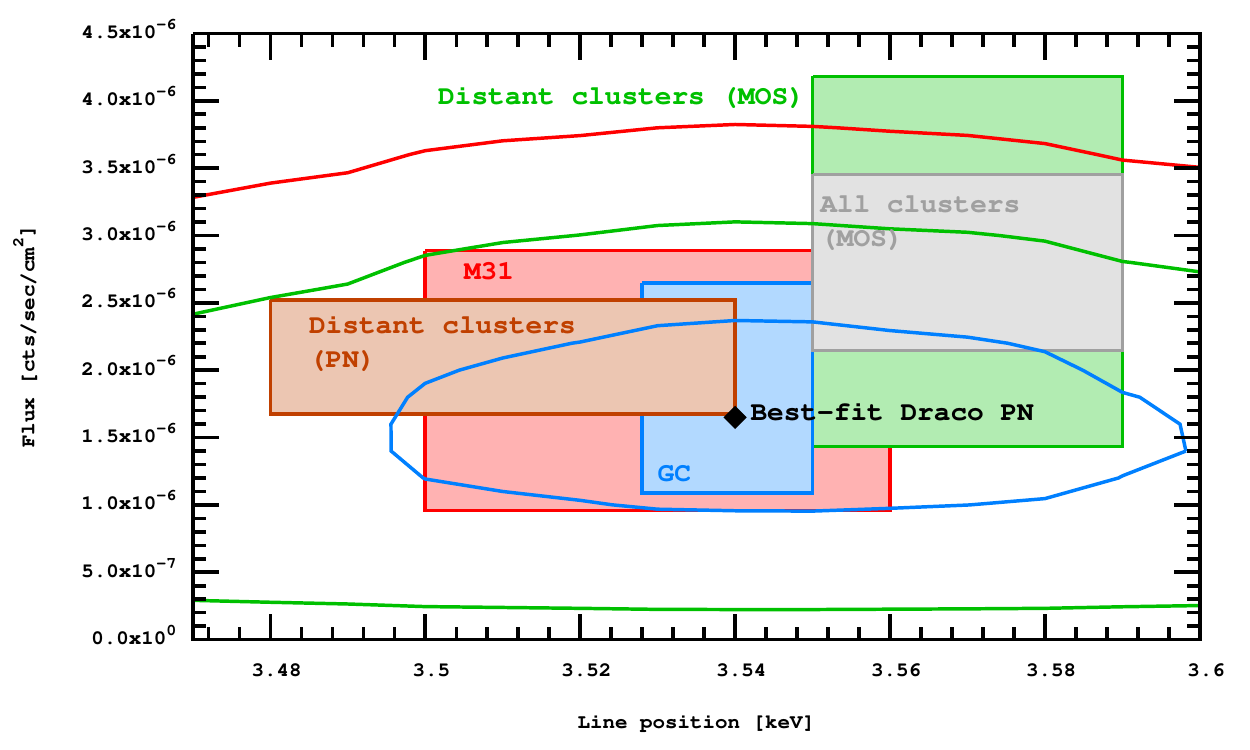}
  \caption{\textit{Left:} Common fit to the MOS1, MOS2, and PN cameras (see
    text for details). The normalization of the $3.5$~keV line between cameras
    is fixed according to the $\Omega_\fov$ ratios (see
    Table~\protect\ref{tab:observations}). Filled rectangles
    show the range of fluxes predicted from previous works. The sizes of the
    regions take into account $\pm 1\sigma$ errors on the measured line fluxes and
    positions. The height of the rectangles also reflects additional spread in expected DM
    signals from the specified objects. The previous bounds are based on:
    \protect\cite{Bulbul:14a} (``All clusters'' and ``Distant clusters''
    samples), \protect\cite{Boyarsky:14a} (``M31'') and \protect\cite{Boyarsky:14b}
    (``GC''). In particular, for ``All clusters'' and ``Distant clusters'' samples, we
    included an additional 20\% uncertainty on its expected DM signal compared
    to the average values shown in Table~5
    of~\protect\cite{Bulbul:14a}, see Sec.~4.1.2 of~\protect\cite{Vikhlinin:08} for 
    detailed discussion. 
    \textit{Right:} same as the left panel, but showing the Draco
    best fit for the PN camera only and the corresponding $\Delta \chi^2 $
    contours (note the different $x$ and $y$ ranges). }
  \label{fig:results_comparison}
\end{figure*}

\section{Discussion}
\label{sec:predictions_M31}

We analysed 26 observations of Draco dSph performed with the \xmm\ during its
AO14 programme. We find a $2.3\sigma$ significant positive line-like residual
at $E=3.54 \pm 0.06$~keV in the PN spectrum. A positive $1\sigma$ residual is also seen at the
position $E=3.60\pm 0.07$~keV with the flux
$F_{\mos2} = (0.76\pm 0.66)\times 10^{-6}\flux$.  Their centroids are
within $1\sigma$ as the right panel of Fig.~\ref{fig:mos2-line-contours}
illustrates. 
 The MOS1 camera had the lowest statistics due to the loss of two
CCDs. It does not show the line but the absence of the signal is consistent with PN and MOS2 at
$1\sigma$ level. The common fit of MOS1, MOS2 and PN camera performed at Sec.~\ref{sec:common-fit}
does not show the presence of a significant positive residual at $\sim$3.5~keV. 
As it is unclear how the common fit is
affected by the uncertainties of cross-calibration between the three cameras, and because we are 
conservatively interested in this paper in exclusion rather than detection, we will use as our main result 
the $2\sigma$ upper bound from the common fit
$F_{2\sigma} < 2.9\times 10^{-9}\flux/\unit{arcmin^2}$, which approximately coincides with
the best-fit PN flux, see Fig.~\ref{fig:results_comparison}.

\subsection{Implications for dark matter decay lifetime}

If one interprets this signal as a line from the dark matter decay, its
flux is related to the dark matter particle's lifetime $\tau_{\dm}$ via
\begin{equation}
  F=\frac{ M_\text{fov}}{4\pi D_L^2}\frac1{\tau_{\dm} m_{\dm}} =
  \frac{\Omega_\text{fov}}{4\pi} \S_\dm \frac1{\tau_{\dm} m_{\dm}} 
\end{equation}
where $m_{\dm}$ is the DM particle mass (equal to $2\times E_\text{line}$),
$M_\text{fov}$ is the DM mass in the field-of-view of the camera, 
$D_L$ is the luminosity distance, and in the second equality we
introduced the \emph{average DM column density}, $\S_\dm$ within the FoV
$\Omega_\text{fov}$.

The expected DM signal from Draco dSph is estimated based on the most recent
stellar kinematics data, modeled in~\cite{Geringer-Sameth:14}. The average
column density of Draco within the central $14'$ is
$\S_\draco = \unit[168]{M_\odot/pc^2}$ (with a typical error of
$\sim20\%$~\citep{Geringer-Sameth:14}).  An additional contribution from the
Milky Way halo in the direction of the Draco dSph was adopted at the level of
$\S_\text{MW} = \unit[93]{M_\odot/pc^2}$ -- based on the profile
of~\cite{Weber:09}.  The scatter in the values of the MW column density ranges
from $\unit[56]{M_\odot/pc^2}$~\citep{Xue:08} to
$\unit[141]{M_\odot/pc^2}$~\citep{Nesti:13}. The resulting column density we
adopt in the direction of Draco is $\S_\draco = \unit[261^{+82}_{-65}]{M_\odot/pc^2}$.

The corresponding lifetime, inferred from the Draco PN camera observation is
$\tau_\draco = (5.1-21.9)\times 10^{27}\unit{sec}$ depending on dark matter column 
densities in the direction of 
Draco:
\begin{equation}
  \label{eq:3}
  \tau_\draco = \left\{
    \begin{aligned}
      9.6^{+7.1}_{-2.8}\times 10^{27}\unit{sec} && \S_\draco = (168+93)\unit{M_\odot/pc^2} \\
      7.2^{+5.3}_{-2.1}\times 10^{27}\unit{sec} && \S_\draco = (140+56)\unit{M_\odot/pc^2}\\
     12.6^{+9.3}_{-3.7}\times 10^{27}\unit{sec} && \S_\draco = (202+141)\unit{M_\odot/pc^2}
    \end{aligned}\right.
\end{equation}

\subsection{Comparison with the previous studies of the 3.5~keV line}

Is this lifetime of this line compatible with the previous observations of
$3.5$~keV line? The answer is
affirmative (see Fig.~\ref{fig:results_comparison}). 
The comparison with the previous  detection in the central $14'$ of the Andromeda galaxy (M31)
depends on both uncertainty of the flux measurement and
the column density in direction of the M31 central part, $\S_{\rm M31}$.
The DM column density in the central
$14'$ has been estimated in~\cite{Boyarsky:07a,Boyarsky:10a,Boyarsky:14a} and
references therein.  In this work, we adopt two values of $\S_{\rm M31}$:
$\S_\text{M31,med} = \unit[600]{M_\odot/pc^2}$ (based on profile from~\cite{Widrow:05}) and
$\S_\text{M31,max} = \unit[1000]{M_\odot/pc^2}$ (based
on~\cite{Geehan:06,Tempel:07}, see~\cite{Boyarsky:07a,Boyarsky:10a} for the
discussion of various profiles). The typical errors on M31 column densities
are at the order of 50\%.
Notice, that a maximal disk large core profile of \cite{Corbelli:09} having
$\S\approx \unit[120]{M_\odot/pc^2}$ would be
incompatible with DM interpretation of the signal. The
predicted lifetime would be too short to be consistent with the strength of
the signal. This reiterates the conclusion, already made
in~\cite{Boyarsky:14b} -- the dark matter interpretation of the $3.5$~keV line
holds only if the density profiles of spiral galaxies (M31 and Milky Way) are
cuspy.

\begin{table*}
\scriptsize
  \centering
  \begin{tabular}{lccc|c|c|c}
\toprule 
   Object/  & Observed flux         & $\S_\dm$          & $\tau_\dm$ & \multicolumn{3}{c}{Predicted
                                                                     flux
                                                                      $[10^{-6}\flux]$}
    \\
    Camera       & $[10^{-6} \flux]  $   & $[M_\odot/\unit{pc}^2]$ & $[10^{27}~\unit{sec}]$  & MOS1                                                & MOS2                   & PN                     \\
\otoprule
& & 600                     & $7.3^{+2.6}_{-1.8}$       & $1.28^{+0.42}_{-0.34}$                              & $2.31^{+0.76}_{-0.61}$ & $2.20^{+0.72}_{-0.58}$ \\
M31 (MOS)  & $4.9^{+1.6}_{-1.3}$ &\cline{1-5}\\
&   & 1000                    & $12.2^{+4.4}_{-3.0}$      & $0.77^{+0.25}_{-0.20}$                              & $1.38^{+0.45}_{-0.37}$ & $1.32^{+0.43}_{-0.35}$ \\
\midrule
73 (all) stacked clusters (MOS)  & $4.0^{+0.8}_{-0.8}$ & 430 &  $5.7^{+1.7}_{-1.5}$ & $1.65^{+0.59}_{-0.38}$ & $2.98^{+1.06}_{-0.68}$ & $2.82^{+1.01}_{-0.65}$ \\
\midrule
69 (distant) stacked clusters (MOS) & $2.1^{+0.4}_{-0.5}$ & 255  & $6.8^{+2.1}_{-2.1}$ & $1.38^{+0.43}_{-0.43}$ & $2.50^{+0.77}_{-0.77}$ & $2.36^{+0.73}_{-0.73}$ \\
\midrule
Perseus with core (MOS)  & $52.0^{+24.1}_{-15.2}$ & 682 & $0.75^{+0.23}_{-0.27}$ & $12.5^{+7.0}_{-2.9}$ & $22.7^{+12.8}_{-5.3}$ & $21.4^{+12.0}_{-5.0}$ \\ 
\midrule
GC (MOS)  & $24^{+12}_{-11}$        & 3370
                                                                    \citep{Smith:06}
                                                        & $7.9^{+6.7}_{-2.6}$
                                                                     &
                                                                       $1.19^{+0.58}_{-0.55}$ & $2.15^{+1.05}_{-0.99}$ & $2.03^{+1.00}_{-0.93}$ \\
\bottomrule
  \end{tabular}
  \caption{Predicted line flux based on~\protect\citet{Boyarsky:14b,Boyarsky:14a,Bulbul:14a}. The value for the Draco-deduced
    lifetime, $\tau_\draco$ is show in Eq.~\protect\eqref{eq:3}.}
  \label{tab:predicted_flux}
\end{table*}

The DM distributions in Galactic Center region has been summarized
in~\cite{Boyarsky:14b}.  As a DM column density proxy, we used the
distribution of \cite{Smith:06} that gives
$\S_\text{GC} = \unit[3370]{M_\odot/pc^2}$ The spread of allowed values of
column density for the Galactic Center distributions
is larger than the order of magnitude, see Fig.~2 of~\cite{Boyarsky:14b}.

\cite{Lovell:14} analysed Milky Way-like halo in Aquarius simulation,
identified there DM halos that could be considered as hosting Draco
dSph. It was found that the best agreement between the simulations and
observational constraints is for $\tau \sim (6-10)\times 10^{27}$~sec (with
which  the Draco-deduced lifetime~\eqref{eq:3} is fully consistent). They also
predicted the ratio of fluxes $F_\draco /F_{\text{GC}}$ to peak at $0.09$ with
the scatter ranging from $0.04$ to $0.2$ ($95\%$ range). Again, the ratio of the fluxes ($\sim 0.07$)
inferred in this paper based on Draco PN data
is close to the most probably value predicted by~\cite{Lovell:14}.

The dark matter column density of the combined sample of galaxy clusters is
given by Table~5 of~\cite{Bulbul:14a}.  For the sample of ``all distant
clusters'' considered here, the mean value of DM column density is
$\S_\text{clusters} = \unit[255]{M_\odot/pc^2}$ and we assign an additional
$20\%$ error to this data according to Sec.~4.1.2 of~\cite{Vikhlinin:08} (notice that the relative errorbars on central column densities
are at the order of $\sim 2$, see e.g. Table~I of~\cite{Iakubovskyi:15b}). The resulting
lifetime is again listed in Table~\ref{tab:predicted_flux} and is consistent with our measurements.

Finally, a number of works~\citep{Bulbul:14a,Urban:14,Boyarsky:14b,Tamura:14}
already observed that the line from the central region of Perseus galaxy
cluster is too strong to be compatible with other detections.  This
signal can only be reconciled with the simple decaying DM hypothesis either if there is a
strong additional emission from the atomic lines (e.g., Ar~{\sc xvii} satellite
line~\citep{Bulbul:14a}) in the central region, or if there is a clump of dark
matter, making the central column density much larger than estimated based on
the temperature profiles~\citep{Bulbul:14a}. The forthcoming \emph{Astro-H}
mission~\citep{Kitayama:14,Mitsuda:14,Takahashi:12} will be able to resolve the
issue with the origin of the emission from the Perseus center, both in core
and outskirts, as well as in other bright clusters.  
On the other hand, outskirts of the Perseus cluster, considered
in~\cite{Boyarsky:14a} are compatible with the decaying DM interpretation for
lifetimes up to $8\times 10^{27}$~sec, compatible with Draco-deduced lifetime
within $1\sigma$.

\subsection{Comparison with another recent analysis of Draco extended dataset}

In the recent paper~\citet{Jeltema:15} (JP15 in what follows),
different results of the analysis of the same Draco dSph data have been reported 
claiming that the dark matter interpretation of the 3.5~keV line is excluded at 99\% CL. 
It is difficult to explain the discrepancies with our results without knowing
the details of their data analysis. But we expect that the combination of the
following factors may be important here.
\begin{inparaenum}[\it (i)]\item 
The continuum model of JP15 does not include the extragalactic
  \texttt{powerlaw} component, which affects the shape of the continuum in the
  3--4~keV range at the level of a few \% in a non-trivial,
  non-monotonic way due to the energy dependence of the effective area.
\item The lines at $\sim 3.3\kev$ and $3.7\kev$ are detected but unmodeled in
  JP15.  Again, this increases the best-fit continuum level in the energy
  range 3--4 keV, which would artificially strengthen the upper bound on a 3.5
  keV line.
\item JP15 give the highest weight to their more stringent MOS upper
limit. As we see in our spectra, the PN camera shows a positive $\sim 2\sigma$ 
residual at the expected line energy. When searching for weak
signals at or below the telescope sensitivity, it is statistically
proper to combine resutls from the independent detectors (provided
they are mutually consistent), as we do. There is no reason to neglect
the PN constraint -- especially since in this case it is the most
sensitive camera of the three, even with the shortest clean exposure.
\end{inparaenum}

\section*{Acknowledgements.} 

We would like to thank K.~Abazajian, G.~Bertone, A.~Geringer-Sameth, M.~Lovell, 
M.~Walker, C.~Weniger for collaboration, discussion and useful comments.
E.~B.\ acknowledges support by NASA through grant NNX123AE77G.
The work of D.~I.\ has been partially supported from the Swiss National Science Foundation 
grant SCOPE IZ7370-152581, the grant No.~F64/42-2015 of the State Fund for Fundamental Research of Ukraine, 
the Program of Cosmic Research of the National Academy of Sciences of Ukraine, 
and the State Programme of Implementation of Grid Technology in Ukraine. The
Draco dSph observations were performed as a part of AO-14 Very Large Programme obtained with XMM-Newton, an ESA science mission with instruments and contributions directly funded by ESA Member States and NASA.

\appendix

\section{Data preparation and analysis}
\label{sec:data-analysis}

We analysed observations of Draco dSph performed in 2015 by the European
Photon Imaging Cameras (EPIC)~\citep{Strueder:01,Turner:00} on-board of the
X-ray Multi-Mirror Observatory \xmm~\citep{Jansen:01} as a part of AO-14
campaign (proposal 76480, PI: A.~Boyarsky).  
The 1.4~Ms total requested exposure was divided into 26
observations (Table~\ref{tab:observations}). We processed
these observations from raw data using Extended Source Analysis
Software (ESAS)~\citep{Kuntz:08} provided as part of the  \xmm\ Science Analysis System 
SAS v.14.0.0, with calibration files current as of December 1, 2015.
Time intervals affected by soft proton flares were
rejected using ESAS procedure \texttt{mos-filter} with standard spectral
cuts. This rejection removed $\sim $30\% ($\sim $50\%) of raw exposure for
MOS(PN) cameras, comparable to other methods
described in e.g. Sec.~8.4.1 of~\cite{Iakubovskyi:13}. 
While the exposure reduction is significant, our tests showed that the sensitivity to the line does not 
increase if we use less-rigorous cleaning, because of the resulting increase of the continuum brightness.
We excised the unrelated X-ray point sources using the ESAS
procedure {\tt cheese}, which masks the sky regions \emph{around the detected point sources} 
of the $\geq$36" radius which corresponds to the removal of $\geq$70\% per cent of total 
encircled energy. 
For each
observation, exposures and fields-of-view after removal of proton flares and
point sources are listed in Table~\ref{tab:observations}. Spectra and response matrices from MOS and PN
cameras produced by ESAS procedure \texttt{mos-spectra} were then combined
using FTOOL \texttt{addspec} and binned with FTOOL \texttt{grppha} by 65~eV 
to have excellent statistics ($\sim 1-2$\% rms
variation per bin obtained from $\gtrsim 2000$ counts per bin in our spectra 
-- comparable to the expected line excess above the
continuum) while still resolving the spectral lines.
For PN camera, we
additionally corrected the obtained spectra for out-of-time events using the
standard procedure.

We stress that for our current purpose, given
the low expected flux of the spectral line in question and the limited
spectral resolution of the CCD detectors, accurate modeling of the
continuum emission in the immediate vicinity of the line is crucial.
For this, we must account for all the faint detector lines and
features that may bias the continuum even at a percent level. We do this by
creating a spectral model that includes the detector and sky X-ray
background components as described below.
We analyse the spectra in the band 2.8--10~keV (for MOS spectra) and 2.8--6.8 +
10--11~keV for PN spectra (to stay well away from the mirror edges found at $E\lesssim 2.5$ keV) 
using standard X-ray spectral fitting tool
\texttt{Xspec}. The region 6.8-10~keV was removed from PN
camera to avoid modeling very strong instrumental lines (Ni~K$\alpha$, Cu~K$\alpha$, Ni~K$\beta$ and 
Zn~K$\alpha$),
see Table~\ref{tab:model-parameters} that cannot
be modeled adequately with simple Gaussian profiles at such good
statistics. However, adding the high
energy bins (at 10-11~keV) allows us to constrain the slope of the instrumental
continuum.

The instrumental background is modeled by a sum of unfolded broken
\texttt{powerlaw} continuum and several narrow \texttt{gaussians}
corresponding to bright fluorescent lines originated inside the instrument.
Because astrophysical emission from dwarf spheroidal galaxies is
negligible~\citep{Gizis:93,Boyarsky:06d,Jeltema:08,Riemer-Sorensen:09,Boyarsky:10a,Sonbas:15},
astrophysical model represents contribution of Cosmic X-ray background (CXB)
modeled by a folded \texttt{powerlaw} continuum. The best-fit values of the flux and powerlaw index
of Cosmic X-ray background were allowed to
change within 95\% CL to the best-fit values from \cite{Moretti:09}.  
Neutral hydrogen absorption column
density was fixed to the weighted value
n$_\text{H} = 2.25 \times 10^{20}$~cm$^{-2}$ obtained from
Leiden-Argentine-Bonn survey~\citep{Kalberla:05} in direction of Draco.
The obtained CXB parameters from MOS1, MOS2 and PN are consistent with each other and with those 
summarized in~\cite{Moretti:09} 
at $<90$~\% level. Moreover, if one freezes the CXB parameters in PN camera at the level of 
MOS1/MOS2 or at the level of~\cite{Moretti:09} the best-fit normalization of 3.5~keV 
line in PN camera changes by no more than 10\%, much beyond its statistical errors ($\sim$40~\%, according 
to Eq.~(\ref{eq:2})). 
We conclude therefore that the origin of the positive line-like residual at 3.5~keV 
seen in PN camera is not due to variation of CXB parameters from its `conventional' level.

\begin{table*}
\centering
\begin{tabular}[c]{lccc}
\hline
Parameter                                       & MOS1  & MOS2  & PN      \\
\hline
CXB \texttt{powerlaw} index           & 1.40$^{+0.32}_{-0.06}$ & 1.36$^{+0.07}_{-0.08}$  & 1.61$^{+0.13}_{-0.06}$    \\
CXB \texttt{powerlaw} flux            & 1.37$^{+0.34}_{-0.67}$ &
                                                                           1.40$^{+0.30}_{-0.40}$  & 2.03$^{+0.38}_{-0.30}$    \\
at 2-10~keV [10$^{-11}$ erg/sec/cm$^2$/deg$^2$] &       &       &         \\
Instrumental background \texttt{powerlaw} index            & 0.31$^{+0.06}_{-0.05}$  & 0.26$^{+0.05}_{-0.04}$  & 0.36$^{+0.03}_{-0.02}$    \\
Extra gaussian line position at $\sim$3.0~keV [keV]  & ---   & 3.045$^{+0.032}_{-0.031}$ & --- \\
Extra gaussian line flux at $\sim$3.0~keV [$10^{-3}$~cts/s] & --- & 0.3$^{+0.1}_{-0.2}$ & --- \\
K~K$\alpha$ line position [keV]             & 3.295$^{+0.039}_{-0.046}$ & 3.380$^{+0.025}_{-0.020}$ & ---     \\
K~K$\alpha$ line flux [$10^{-3}$~cts/s]          & 0.3$^{+0.1}_{-0.2}$ & 0.5$^{+0.1}_{-0.2}$ & ---     \\
Ca~K$\alpha$ line position [keV]            & ---   & 3.770$^{+0.017}_{-0.023}$ & ---     \\
Ca~K$\alpha$ line flux [$10^{-3}$~cts/s]      & ---   & 0.5$^{+0.2}_{-0.1}$ & ---     \\
Ti~K$\alpha$ line position [keV]                & 4.530$^{+0.028}_{-0.028}$ & ---   & 4.530$^{+0.002}_{-0.014}$   \\
Ti~K$\alpha$ line flux [$10^{-3}$~cts/s]        & 0.4$^{+0.1}_{-0.1}$ & ---   & 4.6$^{+0.4}_{-0.3}$   \\
Cr~K$\alpha$ line position [keV]                & 5.420$^{+0.005}_{-0.005}$ & 5.431$^{+0.004}_{-0.005}$ & 5.445$^{+0.001}_{-0.001}$   \\
Cr~K$\alpha$ line flux [$10^{-3}$~cts/s]         & 2.1$^{+0.1}_{-0.2}$ & 3.7$^{+0.2}_{-0.1}$ & 11.2$^{+0.4}_{-0.3}$   \\
Mn~K$\alpha$ line position [keV]                & 5.919$^{+0.006}_{-0.008}$ & 5.901$^{+0.005}_{-0.005}$ & 5.910$^{+0.014}_{-0.015}$   \\
Mn~K$\alpha$ line flux [$10^{-3}$~cts/s]         & 1.8$^{+0.1}_{-0.1}$ & 3.2$^{+0.2}_{-0.1}$ & 3.3$^{+0.4}_{-0.3}$   \\
Fe~K$\alpha$ line position [keV]                & 6.386$^{+0.020}_{-0.025}$ & 6.424$^{+0.006}_{-0.018}$ & 6.395$^{+0.014}_{-0.013}$   \\
Fe~K$\alpha$ line flux [$10^{-3}$~cts/s]         & 1.6$^{+0.4}_{-0.5}$ & 3.7$^{+0.2}_{-1.0}$ & 9.7$^{+1.1}_{-1.0}$   \\
Fe~K$\beta$ line position [keV]                 & 7.128$^{+0.038}_{-0.025}$ & 7.148$^{+0.037}_{-0.038}$ & n/i$^a$ \\
Fe~K$\beta$ line flux [$10^{-3}$~cts/s]         & 0.5$^{+0.1}_{-0.1}$ & 0.4$^{+0.1}_{-0.2}$ & n/i$^a$ \\
Ni~K$\alpha$ line position [keV]                & 7.460$^{+0.016}_{-0.015}$ & 7.479$^{+0.006}_{-0.008}$ & n/i$^a$ \\
Ni~K$\alpha$ line flux [$10^{-3}$~cts/s]        & 0.8$^{+0.2}_{-0.1}$ & 2.2$^{+0.1}_{-0.2}$ & n/i$^a$ \\
Cu~K$\alpha$ line position [keV]                & 8.021$^{+0.029}_{-0.026}$ & 8.045$^{+0.015}_{-0.025}$ & n/i$^a$ \\
Cu~K$\alpha$ line flux [$10^{-3}$~cts/s]        & 0.5$^{+0.1}_{-0.1}$ & 0.9$^{+0.2}_{-0.1}$ & n/i$^a$ \\
Ni~K$\beta$ line position [keV]                 & 8.251$^{+0.039}_{-0.037}$ & 8.267$^{+0.042}_{-0.052}$ & n/i$^a$ \\
Ni~K$\beta$ line flux [$10^{-3}$~cts/s]         & 0.4$^{+0.1}_{-0.2}$ & 0.4$^{+0.1}_{-0.2}$ & n/i$^a$ \\
Zn~K$\alpha$ line position [keV]                & 8.620$^{+0.023}_{-0.016}$ & 8.629$^{+0.021}_{-0.015}$ & n/i$^a$ \\
Zn~K$\alpha$ line flux [$10^{-3}$~cts/s]         & 0.8$^{+0.1}_{-0.1}$ & 0.9$^{+0.2}_{-0.1}$ & n/i$^a$ \\
Au~L$\alpha$ line position [keV]                & 9.716$^{+0.007}_{-0.005}$ & 9.710$^{+0.009}_{-0.004}$ & n/i$^a$ \\
Au~L$\alpha$ line flux [$10^{-3}$~cts/s]         & 3.0$^{+0.2}_{-0.2}$ & 3.4$^{+0.1}_{-0.2}$ & n/i$^a$ \\
\hline
Overall quality of fit ($\chi^2$/dof)        & 73.8/86 & 79.3/75 & 47.0/58 \\

\end{tabular}
\caption{Model parameters for MOS1, MOS2 and PN cameras including positions and fluxes of instrumental 
line candidates. 
\newline
$^a$ In PN, bright instrumental lines at 7-10~keV are not 
included (n/i) to our model due to large residuals appearing when these lines are modeled. 
Instead, we included high-energy range above 10~keV to our PN to 
improve the instrumental continuum modeling.}
\label{tab:model-parameters}
\end{table*}

To check for possible instrumental gain variations, 
we splitted our datasets into three smaller subsets grouped by observation time, see
observations 1-9, 10-17, and 18-26 in Table~\ref{tab:observations}. 
For each dataset, we presented average position
for Cr~K$\alpha$ and Mn~K$\alpha$ instrumental lines, see Table~\ref{tab:gain-variation} for details. No 
systematic gain variations across the datasets is detected.

\begin{table*}
\centering
\begin{tabular}[c]{lcc}
\hline
Parameter          & Cr~K$\alpha$ position [keV]  & Mn~K$\alpha$ position [keV] \\
\hline
MOS1, obs. 1-9 & 5.427$^{+0.013}_{-0.012}$ & 5.922$^{+0.013}_{-0.013}$  \\
MOS1, obs. 10-17 & 5.416$^{+0.009}_{-0.006}$ & 5.914$^{+0.011}_{-0.010}$  \\
MOS1, obs. 18-26 & 5.402$^{+0.014}_{-0.011}$ & 5.930$^{+0.031}_{-0.031}$  \\
MOS1, full dataset & 5.420$^{+0.005}_{-0.005}$ & 5.919$^{+0.006}_{-0.008}$  \\
\hline
MOS2, obs. 1-9 & 5.436$^{+0.005}_{-0.007}$ & 5.895$^{+0.009}_{-0.009}$  \\
MOS2, obs. 10-17 & 5.431$^{+0.005}_{-0.006}$ & 5.918$^{+0.008}_{-0.007}$  \\
MOS2, obs. 18-26 & 5.431$^{+0.008}_{-0.007}$ & 5.888$^{+0.008}_{-0.008}$  \\
MOS2, full dataset & 5.431$^{+0.004}_{-0.005}$ & 5.901$^{+0.005}_{-0.005}$  \\
\hline
PN, obs. 1-9 & 5.443$^{+0.002}_{-0.013}$ & 5.916$^{+0.042}_{-0.036}$  \\
PN, obs. 10-17 & 5.431$^{+0.014}_{-0.001}$ & 5.955$^{+0.059}_{-0.059}$  \\
PN, obs. 18-26 & 5.446$^{+0.013}_{-0.001}$ & 5.909$^{+0.016}_{-0.016}$  \\
PN, full dataset & 5.445$^{+0.001}_{-0.001}$ & 5.910$^{+0.014}_{-0.015}$  \\
\hline
\end{tabular}
\caption{Best-fit positions and 1$\sigma$ errors for Cr~K$\alpha$ and Mn~K$\alpha$ instrumental lines
detected in three different subsets of our dataset grouped by the time of observation, see observations 1-9, 10-17 and
18-26 from Table~\protect\ref{tab:observations}. The lines position listed for our full dataset also listed
in Table~\protect\ref{tab:model-parameters} is shown for comparison.
No systematic gain variations across the datasets is detected.}
\label{tab:gain-variation}
\end{table*}


\let\jnlstyle=\rm\def\jref#1{{\jnlstyle#1}}\def\aj{\jref{AJ}}
  \def\araa{\jref{ARA\&A}} \def\apj{\jref{ApJ}\ } \def\apjl{\jref{ApJ}\ }
  \def\apjs{\jref{ApJS}} \def\ao{\jref{Appl.~Opt.}} \def\apss{\jref{Ap\&SS}}
  \def\aap{\jref{A\&A}} \def\aapr{\jref{A\&A~Rev.}} \def\aaps{\jref{A\&AS}}
  \def\azh{\jref{AZh}} \def\baas{\jref{BAAS}} \def\jrasc{\jref{JRASC}}
  \def\memras{\jref{MmRAS}} \def\mnras{\jref{MNRAS}\ }
  \def\pra{\jref{Phys.~Rev.~A}\ } \def\prb{\jref{Phys.~Rev.~B}\ }
  \def\prc{\jref{Phys.~Rev.~C}\ } \def\prd{\jref{Phys.~Rev.~D}\ }
  \def\pre{\jref{Phys.~Rev.~E}} \def\prl{\jref{Phys.~Rev.~Lett.}}
  \def\pasp{\jref{PASP}} \def\pasj{\jref{PASJ}} \def\qjras{\jref{QJRAS}}
  \def\skytel{\jref{S\&T}} \def\solphys{\jref{Sol.~Phys.}}
  \def\sovast{\jref{Soviet~Ast.}} \def\ssr{\jref{Space~Sci.~Rev.}}
  \def\zap{\jref{ZAp}} \def\nat{\jref{Nature}\ } \def\iaucirc{\jref{IAU~Circ.}}
  \def\aplett{\jref{Astrophys.~Lett.}}
  \def\apspr{\jref{Astrophys.~Space~Phys.~Res.}}
  \def\bain{\jref{Bull.~Astron.~Inst.~Netherlands}}
  \def\fcp{\jref{Fund.~Cosmic~Phys.}} \def\gca{\jref{Geochim.~Cosmochim.~Acta}}
  \def\grl{\jref{Geophys.~Res.~Lett.}} \def\jcp{\jref{J.~Chem.~Phys.}}
  \def\jgr{\jref{J.~Geophys.~Res.}}
  \def\jqsrt{\jref{J.~Quant.~Spec.~Radiat.~Transf.}}
  \def\memsai{\jref{Mem.~Soc.~Astron.~Italiana}}
  \def\nphysa{\jref{Nucl.~Phys.~A}} \def\physrep{\jref{Phys.~Rep.}}
  \def\physscr{\jref{Phys.~Scr}} \def\planss{\jref{Planet.~Space~Sci.}}
  \def\procspie{\jref{Proc.~SPIE}} \let\astap=\aap \let\apjlett=\apjl
  \let\apjsupp=\apjs \let\applopt=\ao \def\jcap{\jref{JCAP}}

\bsp	
\label{lastpage}
\end{document}